\documentclass[reprint,aps,prd,superscriptaddress,floatfix,nofootinbib,amsmath,amssymb,showkeys,showpacs]{revtex4-1}
\usepackage{graphicx,color,dsfont,bm}
\usepackage[utf8]{inputenc}
\usepackage{float}
\usepackage{longtable}
\begin{document}
\title{Theoretical constraints on scalar parameters in the compact 341 model}
\author{Meriem \surname{Djouala}}
\email{djouala.meriem@umc.edu.dz}
\author{Noureddine \surname{Mebarki}}
\email{nnmebrki@yahoo.fr}
	\affiliation{Laboratoire de physique mathématique et subatomique. Physics Department, University of Mentouri,	Constantine 1, Constantine P.O. Box, 325 Ain El Bey Way, 25017 Constantine, Algeria}	
\begin{abstract}
Theoretical constraints on the scalar potential of the compact 341 model with three quadruplets scalar fields are discussed. It is shown that, in order to ensure good behavior of the potential and the viability of the model, the criteria such as copositivity, minimization, perturbative unitarity, perturbativity of the scalar couplings and no ghost scalar bosons (scalar bosons masses positivity) are imposed and bounds on the scalar couplings are obtained. Moreover, the existence of the Landau pole in the model imposes stringent limits.
\end{abstract}
\maketitle
\section{Introduction}
\label{sec:Intro}
~~Despite of the success of the Standard Model some fundamental questions such as matter-anti matter asymmetry, CP violation, dark matter ect....remain unsolved. Thus, going beyond the Standard Model becomes mandatory. Among the interesting proposed theories beyond the standard model (BSM), the so called 341 models  based on the Lie gauge group $SU(3)_{C}\otimes SU(4)_{L}\otimes$ $U(1)_{X}$ \cite{ref1,ref9,ref444}. In the literature, there are many classifications of this model depending on the existence or not of fermions with exotic charges, structure of the scalar potential and spontaneous symmetry breaking of the gauge group. These models are usually parameterized by two parameters $\beta$ and $\gamma$ \cite{ref1, ref9,ref10}.\\
\indent In this paper, we focus on the compact 341 model with exotic electric charges where $\beta=\frac{-1}{\sqrt{3}}$ and $\gamma=\frac{-4}{\sqrt{6}}$ \cite{ref1}. The most attractive feature of this model is the fact that the two chiralities of the lightest leptons are part of the SU(4) gauge group fundamental representation, unifying each leptons family in a single multiplet. Moreover, the existence of the three families replication of leptons and quarks in the physical spectrum can be explained as a result of the triangle gauge anomaly cancellation together with the QCD asymptotic freedom.\\
\indent As it was pointed out \cite{ref1}, the gauge anomalies cancellatios requires that the three quarks generations have to belong to different $SU(4)_{L}$ representations:
two left handed  generations $Q_{iL}=(d_{i},u_{i},D_{i},J_{i})_{L}$ (i=2,3) lie in the 4  fundamental representation, whereas, the third one $Q_{1}=(u_{1},d_{1},U_{1},J_{1})_{L}$ together with the three leptons generations $\psi_{l}=(\nu_{l},e_{l},\nu^{c}_{l},e^{c}_{l})$ with different chiralities in the same quadruplet $(l=e,\mu,\tau)$ have to transform under $\bar{4}$ the conjugate fundamental representation (or vice versa). Moreover, the right handed  quarks $u_{jR},d_{jR},J_{jR},D_{kR},U_{1R}$ (k=2,3 and j=1,2,3) transform as singlets. Here  $D_{i}$, $J_{i}$, $U_{1}$, $J_{1}$ and c stand for exotic quarks with charges $\frac{-1}{3}$, $\frac{-4}{3}$, $\frac{2}{3}$ and $\frac{5}{3}$  and charge conjugate respectively. 
Concerning the scalar sector and in order to generate the masses of all particles one has to have three quartets $\chi$, $\rho$ and $\eta$ in the SU(4) fundamental representation. Table \ref{tab:aalo} shows all particles content in the compact 341 model parameterized by $\beta=\frac{-1}{\sqrt{3}}$ and $\gamma=\frac{-4}{\sqrt{6}}$.\\
\indent It is well known that the Standard Model scalar potential has only two scalar couplings. To ensure its stability it is enough to make the Higgs boson quartic coupling positive \cite{ref11,ref13}. However, things are more complicated in BSM models \cite{ref2,ref3,ref43,ref4,ref15,ref43} even at the tree level.\\
\indent The purpose of this paper is to study the theoretical constraints in order to determine the compatible allowed regions of the parameters space.\\
\indent This paper is organized as follows: in section \ref{sec:The model}, we present the basic ingredients and particle content of the compact 341 model. In Section \ref{sec:Constraints on the parameters space}, we discuss the various theoretical constraints on the scalar couplings, including the tree level necessary and sufficient vacuum stability conditions using the orbit space method \cite{ref2} and in order to ensure the boundedness of the scalar potential in any direction of the fields the copositivity criterion is imposed. Other minimization constraints are also discussed. Moreover, the perturbative unitarity, the positivity of the scalar bosons masses and the perturbativity of the scalar potential conditions are derived using the fact that the lightest CP even boson $H_{1}$ is identified to the SM Higgs boson. The combination of all these theoretical constraints, together with the Landau pole limit allow to determine the compatible region of the parameters space. Finally, in section \ref{sec:Conclusion} we draw our conclusions.
\section{The model}
\label{sec:The model}
The compact 341 model is based on the gauge group $SU(3)_{C}\otimes SU(4)_{L}\otimes U(1)_{X}$, where, C, L and X stand respectively for color, left chirality and a new quantum number defining the charge associated to the group $U(1)_X$ which is
written as a linear combination together with the diagonal generators $T_{3}$, $T_{8}$ and $T_{15}$ of the gauge group $SU(4)_{L}\otimes U(1)_{X}$ to determine the electric charge operator:
\begin{align}
Q=T_{3}+\beta T_{8}+\gamma T_{15}+ XI_{4\times4}
\end{align}
where $\beta$ and $\gamma$ are parameters to be fixed according to the field distribution in the quadruplets. In this paper the fermionic assignment is determined with  $\beta=\frac{-1}{\sqrt{3}}$ and $\gamma=\frac{-4}{\sqrt{6}}$ as it is shown in Table \ref{tab:aalo}.
 \begin{table}[H]
 \caption{\label{tab:aalo} The complete anomaly free particle content in the compact 341 model. Where the quantum numbers between parenthesis indicate the respective representation (rep) for the quadruplets (triplets) under the 341 (331) symmetry and F refers to Flavors.}
 \begin{ruledtabular}
 \begin{tabular}{c|c|c|c|c}
 Name & 341 rep& 331 rep & components& F \\
 \hline
 $\psi_{\alpha L}$&(1,4,$0$)&(1,3,$\frac{-1}{3}$)+(1,1,1)&($\nu_{\alpha},l_{\alpha},\nu_{\alpha}^{C},l_{\alpha}^{C}$)&3\\
 $Q_{iL}(i=2,3)$&(3,$\overline{4}$,$\frac{-1}{3})$&(3,$\bar{3}$,0)+(3,1,$\frac{-4}{3}$)&$(d_{i},u_{i},D_{i},J_{i})$&2\\
 $Q_{1L}$&(3,4,$\frac{2}{3})$&(3,3,$\frac{1}{3}$)+(3,1,$\frac{5}{3}$)&$(u_{1},d_{1},U_{1},J_{1})$&1\\
 $u_{jR}(j=1,i)$&($\overline{3}$,1,$\frac{2}{3}$)&($\overline{3}$,1,$\frac{2}{3}$)&$u_{jR}$&4\\
 $d_{jR}(j=1,i)$&($\overline{3}$,1,$\frac{-1}{3}$)&($\overline{3}$,1,$\frac{-1}{3}$)&$d_{jR}$&5\\
 $J_{1R}$&($\overline{3}$,1,$\frac{5}{3}$)&($\overline{3}$,1,$\frac{5}{3}$)&$J_{1R}$&1\\
 $J_{iR}$(i=1,2)&($\overline{3}$,1,$\frac{-4}{3}$)&($\overline{3}$,1,$\frac{-4}{3}$)&$J_{iR}$&2\\
 $\chi$&(1,4,-1)&(1,3,
 $\frac{-4}{3}$)+(1,1,0)&($\chi_{1}^{-}$,$\chi^{--}$,$\chi_{2}^{-}$,$\chi^{0})$&1\\
 $\eta$&(1,4,0)&(1,3,$\frac{-1}{3}$)+(1,1,1)&($\eta_{1}^{0}$,$\eta_{1}^{-}$,$\eta_{2}^{0}$,$\eta_{2}^{+})$&1\\
 $\rho$&(1,4,1)&(1,3,$\frac{2}{3}$)+(1,1,2)&($\rho_{1}^{+}$,$\rho^{0}$,$\rho_{2}^{+}$,$\rho^{++})$&1\\
 \end{tabular}
 \end{ruledtabular}
 \end{table}
In the compact 341 model, the expression of the most general renormalizable and gauge invariant scalar potential is \cite{ref1}:
\begin{align}\label{eq:M}
V(\eta,\rho,\chi)=&\mu_{\eta}^{2}\eta^{\dag}\eta+\mu_{\rho}^{2}\rho^{\dag}\rho+\mu_{\chi}^{2}\chi^{\dag}\chi+\lambda_{1}(\eta^{\dag}\eta)^{2}+\lambda_{2}(\rho^{\dag}\rho)^{2}\nonumber\\&+\lambda_{3}(\chi^{\dag}\chi)^{2}
+\lambda_{4}(\eta^{\dag}\eta)(\rho^{\dag}\rho)+\lambda_{5}(\eta^{\dag}\eta)(\chi^{\dag}\chi)\nonumber\\
&+\lambda_{6}(\rho^{\dag}\rho)(\chi^{\dag}\chi)+\lambda_{7}(\rho^{\dag}\eta)(\eta^{\dag}\rho)+\lambda_{8}(\chi^{\dag}\eta)(\eta^{\dag}\chi)\nonumber\\
&+\lambda_{9}(\rho^{\dag}\chi)(\chi^{\dag}\rho),
\end{align}
Where $\mu^{2}_{\mu\rho\chi}$ are the mass dimension parameters and
$\lambda_{i}$ (i=1..9) are real dimensionless coupling constants.
The spontaneous symmetry breaking (SSB) in the compact 341 model proceeds in three steps:
\begin{align}
\begin{alignedat}{1}
SU(3)_{c}\otimes S&U(4)_{L}\otimes U(1)_{X}\\
&\Downarrow \langle\chi\rangle\\
SU(3)_{c}\otimes S&U(3)_{L}\otimes U(1)_{X'}\\
&\Downarrow \langle\eta\rangle\\
SU(3)_{c}\otimes S&U(2)_{L}\otimes U(1)_{Y}\\
&\Downarrow  \langle\rho\rangle\\
SU(3)_c&\otimes U_{em}
\end{alignedat}
\end{align}
Where after the neutral components of the scalar fields $\eta$, $\chi$ and $\rho$ acquire its vacuum expectation values which are needed to give masses for all the particles in this model, the VEVs structure is represented as: 
$$
\langle \eta \rangle=\frac{1}{\sqrt{2}} \left(
\begin{array}{ccc}
0\\
0 \\
\upsilon_{\eta} \\
0
\end{array}
\right),~ \langle \rho \rangle=\frac{1}{\sqrt{2}} \left(
\begin{array}{ccc}
0\\
\upsilon_{\rho}\\
0 \\
0
\end{array}
\right),~\langle \chi \rangle=\frac{1}{\sqrt{2}}
\left(
\begin{array}{ccc}
0 \\
0 \\
0 \\
\upsilon_{\chi}
\end{array}
\right) 
$$
Where $\upsilon_{\rho}$ is vacuum expectation value (VEV) of the SM. Therefore, the VEVs $\upsilon_{\rho}$, $\upsilon_{\eta}$ and $\upsilon_{\chi}$ require the following hierarchy:
\begin{align}
\upsilon_{\rho}\ll\upsilon_{\eta},\upsilon_{\chi}.
\end{align}
\indent Thus the terms proportional to $\upsilon_{\rho}$ will be neglected comparing to the terms proportional to $\upsilon_{\chi}$ and $\upsilon_{\eta}$. Moreover, in this work and following ref \cite{ref1}, we use $\upsilon_{\eta}=\upsilon_{\chi}$, this relation simplify the masses expressions of all physical particles and all relations in this model. The compact 341 model predicts the existence of new fermions as it is indicated in Table \ref{tab:aalo}. Moreover, its scalar spectrum has three electrically neutral CP even scalars $H_{i}(i=1,2,3)$, four singly charged $H_{1}^{\mp}$ and $H_{2}^{\mp}$ and two doubly charged scalar bosons $H^{\mp\mp}$. Their masses expressions are found from the diagonalization of the $3 \times 3$ and three $2\times 2$ matrices which are written in the basis $(R_{\rho}, R_{\chi}, R_{\eta})$, $(\rho_{2}^{\mp}, \eta_{1}^{\mp})$, $(\eta_{2}^{\mp},\chi_{2}^{\mp})$ and $(\rho^{\mp\mp}, \chi^{\mp\mp})$ respectively:
\begin{align}
M^{2}_{H_{1}}&= \lambda_{2}\upsilon_{\rho}^{2}+\frac{\lambda_{3}\lambda^{2}_{4}+\lambda_{6}(\lambda_{1}\lambda_{6}-\lambda_{4}\lambda_{5})}{\lambda^{2}_{5}-4\lambda_{1}\lambda_{3}}\upsilon_{\rho}^{2},\\
M_{H_{2}}^{2}&= \frac{1}{2}\bigg(\lambda_{1}+\lambda_{3}-\sqrt{(\lambda_{1}-\lambda_{3})^{2}+\lambda_{5}^{2}}
\bigg)\upsilon_{\chi}^{2},\\
M_{H_{3}}^{2}&= \frac{1}{2}\bigg(\lambda_{1}+\lambda_{3}+\sqrt{(\lambda_{1}-\lambda_{3})^{2}+\lambda_{5}^{2}}
\bigg)\upsilon_{\chi}^{2},\\
M^{2}_{H^{\mp}_{1}}&=\frac{\lambda_{7}}{2}\upsilon^{2}_{\eta},\\
M^{2}_{H^{\mp}_{2}}&=\frac{\lambda_{8}}{2}(\upsilon^{2}_{\eta}+\upsilon^{2}_{\chi}),\\
M^{2}_{H^{\mp\mp}}&=\frac{\lambda_{9}}{2}\upsilon^{2}_{\chi}.
\end{align}
With four singly charged $G^{\mp}$, two doubly charged $G^{\mp\mp}$ and five neutral massless Goldstone bosons which will be eaten by the gauge bosons to acquire their masses.\\
From the expressions of the neutral Higgs bosons, one can notice that $M_{H_{1}}\ll M_{H_{2}}\ll M_{H_{3}}$. The lightest neutral scalar $H_{1}$ is assigned to be the SM Higgs boson (we take its mass at 125.09 GeV for our calculations), however, the other heavy scalars $H_{2,3}$ are associated with more heavier scales $\upsilon_{\chi}$ and $\upsilon_{\eta}$. The eigenstates of the last $3 \times 3$ and the three $2\times 2$ matrices lead to the the following fields expressions:
\begin{align}
H_{1}^{\mp}&=\sin\alpha\eta_{1}^{\mp}+\cos\alpha~\rho_{2}^{\mp},\\
H_{2}^{\mp}&=\cos\beta~\eta_{2}^{\mp}+\sin\beta~\chi_{2}^{\mp},\\	
H^{\mp\mp}&=\cos'\alpha~\rho^{\mp\mp}+\sin'\alpha ~\chi^{\mp\mp}, \\
G^{\mp}_{1}&=-\cos\alpha~\eta_{1}^\mp+\sin\alpha~\rho_{2}^\mp,\\
G^{\mp}_{2}&=-\sin\beta~\eta_{2}^\mp+\cos\beta~\chi_{2}^\mp,\\
G^{\mp\mp}&=-\sin'\alpha~\rho^{\mp\mp}+\cos'\alpha~\chi^{\mp\mp}.
\end{align}	
Where:
\begin{align}
\cos\alpha&=\frac{\upsilon_{\eta}}{\sqrt{\upsilon_{\eta}^2+\upsilon_{\rho}^2}},\qquad \sin\alpha=\frac{\upsilon_{\rho}}{\sqrt{\upsilon_{\eta}^2+\upsilon_{\rho}^2}},\\
\cos^{\prime}\alpha&=\frac{\upsilon_{\chi}}{\sqrt{\upsilon_{\chi}^2+\upsilon_{\rho}^2}},\qquad \sin'\alpha=\frac{\upsilon_{\rho}}{\sqrt{\upsilon_{\chi}^2+\upsilon_{\rho}^2}},\\
\cos\beta&=\frac{\upsilon_{\chi}}{\sqrt{\upsilon_{\eta}^2+\upsilon_{\chi}^2}},\qquad \sin\beta=\frac{\upsilon_{\eta}}{\sqrt{\upsilon_{\eta}^2+\upsilon_{\chi}^2}}
\end{align}
In what follows, we set $\cos\alpha$ ($\cos'\alpha$), $\sin\alpha$ ($\sin'\alpha$), $\cos\beta$, $\sin\beta$ as $C_{\alpha} (C'_{\alpha})$, $S_{\alpha} (S'_{\alpha})$, $C_{\beta}$ and $S_{\beta}$ respectively.\\
Furthermore, in our model, besides $W^{\mp}$ and Z, it predicts other new gauge bosons, namely: $K_{1}^{\mp}$, $K^{0} (K^{'0})$, $X^{\mp}$, $V^{\mp\mp}$, $Y^{\mp}$, $Z^{\prime}$ and $Z^{\prime\prime}$. Their masses are given by \cite{ref1}:
\begin{align}
M_{W{\mp}}^{2}&=\frac{g^{2}}{4}\upsilon^{2}_{\rho},~~
M_{K'^{0},K^{0}}^{2}=\frac{g^{2}}{4}\upsilon^{2}_{\eta},~~
M_{K_{1}^{\mp}}^{2}=\frac{g^{2}}{4}\upsilon^{2}_{\eta},\\
M_{X^{\mp}}^{2}&=\frac{g^{2}}{4}\upsilon^{2}_{\chi},~~
M_{V^{\mp\mp}}^{2}=\frac{g^{2}}{4}\upsilon^{2}_{\chi},~~
M_{Z}^{2}=\frac{g^{2}\upsilon_{\rho}^{2}}{4c_{W}^{2}},\\
M_{Y^{\mp}}^{2}&=\frac{g^{2}}{4}(\upsilon^{2}_{\eta}+\upsilon^{2}_{\chi}),~~
M_{Z'}^{2}=\frac{g^{2}c_{W}^{2}\upsilon_{\eta}^{2}}{h_{W}},\\ M_{Z''}^{2}&=\frac{g^{2}\upsilon_{\eta}^{2}\bigg((1-4s_{W}^{2})^{2}+h_{W}^{2}\bigg)}{8h_{W}(1-4s_{W}^{2})}.
\end{align}
Where $h_{W}=3-4S_{W}^{2}$, g represents the gauge coupling of the $SU(4)_{L}$, $s_{ W}$ and $c_{W}$ are the sine and cosine electroweak mixing angle.
\section{Constraints on the parameters space}
\label{sec:Constraints on the parameters space}
\indent The compact 341 model has large
numbers of free scalar parameters. To determine their allowed regions and in order to obtain a viable model, many theoretical constraints have to be imposed on the scalar potential. 
\subsection{Minimization conditions}
\label{sec:Minimization and vacuum stability}
The first set of the theoretical constraints on the scalar parameters comes from the minimization conditions resulted from the first and the second derivative of the scalar potential. They require general conditions to provide the vacuum configuration $\langle \rho \rangle_{0}$, $\langle \chi \rangle_{0}$ and $\langle \eta \rangle_{0}$ to be a minimum of the scalar potential \eqref{eq:M}.\\
 The first derivative $\frac{\partial V}{\partial\phi}|_{\phi=0}=0$ are given by:
\begin{align}\label{eq:the}
\mu_{\eta}^{2}+\lambda_{1}\upsilon_{\eta}^{2}+\frac{1}{2}\lambda_{4}\upsilon_{\rho}^{2}+\frac{1}{2}\lambda_{5}\upsilon_{\chi}^{2}=0,\\
\mu_{\rho}^{2}+\lambda_{2}\upsilon_{\rho}^{2}+\frac{1}{2}\lambda_{4}\upsilon_{\eta}^{2}+\frac{1}{2}\lambda_{6}\upsilon_{\chi}^{2}=0,\label{eq:the2}\\
\mu_{\chi}^{2}+\lambda_{3}\upsilon_{\chi}^{2}+\frac{1}{2}\lambda_{5}\upsilon_{\eta}^{2}+\frac{1}{2}\lambda_{6}\upsilon_{\rho}^{2}=0 \label{eq:the3}.
\end{align}
where we have used $\langle \eta \rangle_{0}=\frac{\upsilon_{\eta}}{\sqrt{2}}$, $\langle \rho \rangle_{0} =\frac{\upsilon_{\rho}}{\sqrt{2}}$ and $\langle \chi \rangle_{0}=\frac{\upsilon_{\chi}}{\sqrt{2}}$. \\
The second derivative test $\frac{\partial ^{2}V}{\partial\phi_{i}\phi_{j}} \bigg|_{\phi=\phi_{0}}$ leads to the so called the Hessian matrix $H_{0}$ evaluated at the vacuum:
$$
H_{0}= \left(
\begin{array}{ccc}
4\lambda_{1}\upsilon_{\eta}^2& 2\lambda_{4}\upsilon_{\eta}\upsilon_{\rho}&2\lambda_{5}\upsilon_{\eta}\upsilon_{\chi} \\
2\lambda_{4}\upsilon_{\eta}\upsilon_{\rho}&4\lambda_{2}\upsilon_{\rho}^2&2\lambda_{6}\upsilon_{\rho}\upsilon_{\chi}\\
2\lambda_{5}\upsilon_{\eta}\upsilon_{\chi}&2\lambda_{6}\upsilon_{\rho}\upsilon_{\chi}&4\lambda_{3}\upsilon_{\chi}^2\\
\end{array}
\right),
$$
where we have used the relations in \eqref{eq:the}, \eqref{eq:the2} and \eqref{eq:the3} in order to simplify the Hessian matrix. Using Sylvester’s criterion and from the positivity of the principal minors, we get the following conditions:
\begin{align}
\lambda_{1}>0, \qquad \lambda_{2}>0, \qquad \lambda_{3}>0,\nonumber\\
-2\sqrt{\lambda_{1}\lambda_{2}}<\lambda_{4}<2\sqrt{\lambda_{1}\lambda_{2}},\nonumber\\
-2\sqrt{\lambda_{1}\lambda_{3}}<\lambda_{5}<2\sqrt{\lambda_{1}\lambda_{3}},\nonumber\\
-2\sqrt{\lambda_{3}\lambda_{2}}<\lambda_{6}<2\sqrt{\lambda_{3}\lambda_{2}},\nonumber\\
\text{det}(H_{0})>0.\label{eq:delenda1}
\end{align}
Without mentioning the expression of $\text{det} (H_0)$, The condition $\text{det} (H_0)>0$ is always true because of the fact that  all the square masses of the CP-even scalars are positive as we will discuss in the next subsection.
\subsection{Boundedness from below}
\label{sec:Boundedness from below}
~~We study the vacuum stability at the tree level, the
conditions which guarantee that the scalar potential is
bounded from below in all directions in the field space as
the field strength approaches infinity. In our case, we face
a more complicated problem even at the tree level since we
have to deal with large numbers of scalar couplings (twelve
couplings). Thus, we introduce a parameterztaion which
greatly reduces the number of variables and make the
problem even more tractable to derive the sufficient and
complete constraints of the potential stability (VS)
where we ignore terms with dimension $d<4$, since
in the limit of large field values, they are negligible in
comparison with the quartic couplings of the scalar po-
tential $V^4(\eta, \rho, \chi)$ [9]:
\begin{align}\label{eq:mimzy}
V^{4}(\eta,\rho,\chi)=&\lambda_{1}(\eta^{\dag}\eta)^{2}+\lambda_{2}(\rho^{\dag}\rho)^{2}+\lambda_{3}(\chi^{\dag}\chi)^{2}
+\lambda_{4}(\eta^{\dag}\eta)(\rho^{\dag}\rho)\nonumber\\&+\lambda_{5}(\eta^{\dag}\eta)(\chi^{\dag}\chi)+\lambda_{6}(\rho^{\dag}\rho)(\chi^{\dag}\chi)+\lambda_{7}(\rho^{\dag}\eta)(\eta^{\dag}\rho)\nonumber\\
&+\lambda_{8}(\chi^{\dag}\eta)(\eta^{\dag}\chi)+\lambda_{9}(\rho^{\dag}\chi)(\chi^{\dag}\rho),
\end{align}
\indent Since we have three different field directions, we define a parametrization of the fields on a sphere: 
\begin{align}
r^2&\equiv \eta^\dag\eta+\rho^\dag\rho+\chi^\dag\chi,\nonumber\\
\eta^\dag\eta&\equiv r^2\cos^2\theta \sin^2\phi,\nonumber\\
\rho^\dag\rho&\equiv r^2\sin^2\theta \sin^2\phi,\nonumber\\
\chi^\dag\chi&\equiv r^2\cos^2\phi,\nonumber\\
\frac{\eta^\dag\rho}{|\eta||\rho|}&\equiv \xi_{1}e^{i\psi_{1}},\nonumber~~  \frac{\eta\rho^\dag}{|\eta||\rho|}\equiv \xi_{1}e^{-i\psi_{1}},\nonumber \\
\frac{\eta^\dag\chi}{|\eta||\chi|}&\equiv \xi_{2}e^{i\psi_{2}},~~\nonumber  \frac{\eta\chi^\dag}{|\eta||\chi|}\equiv \xi_{2}e^{-i\psi_{2}}, \nonumber\\
\frac{\rho^\dag\chi}{|\rho||\chi|}&\equiv \xi_{3}e^{i\psi_{3}},~~ \frac{\rho\chi^\dag}{|\rho||\chi|}\equiv \xi_{3}e^{-i\psi_{3}}.
\end{align}
where we adopt here a parametrization similar to the
one of ref \cite{ref2}, The scalar fields $\eta$, $\chi$ and $\rho$ scan all the fields space, therefore, the radius r scans the domain [0,$\infty$[, the angle $\theta \in$ [0,2$\pi$] and the angle $\phi\in$[0,$\frac{\pi}{2}$], $\xi_{i} (i=1,2,3)\in$ [0,1] \cite{ref2}.\\
\indent Inserting this parameterztaion in the scalar potential \eqref{eq:mimzy}, it is straightforward to write $V^{4}(\rho,\chi,\eta)$ in the following form:  
\begin{align} \label{eq:daad}
V^{4}(r&, \cos^2\theta, \sin^2\theta, \cos^2\phi, \xi_{i} )=r^4\bigg(\lambda_{1}\cos^4\theta\sin^4\phi\nonumber\\&+\lambda_{2}\sin^4\theta\sin^4\phi+\lambda_{3}\cos^4\phi+\lambda_{4}\cos^2\theta\sin^2\theta\sin^4\phi\nonumber\\&+\lambda_{5}\cos^2\theta\cos^2\phi
+\lambda_{6}\sin^2\theta\sin^2\phi\cos^2\phi+\lambda_{7}\xi_{1}^2\nonumber\\&\cos^2\theta\sin^2\phi\sin^4\phi+\lambda_{8}\xi_{2}^2\cos^2\theta\sin^2\phi\cos^4\phi\nonumber\\&+\lambda_{9}\xi_{3}^2\sin^2\theta\cos^2\phi\sin^2\phi
\bigg),
\end{align}
We introduce again the following variables \cite{ref2}:
\begin{align} \label{eq:papa}
x\equiv \cos^2\theta~~~\text{and}~~~y\equiv \sin^2\phi, 
\end{align}
Inserting \eqref{eq:papa} in the expression \eqref{eq:daad}, them we get:
\begin{align}\label{eq:Mr}
V^{4}(r&, \cos^2\theta, \sin^2\theta, \cos^2\phi, \xi_{i} )=y^2\bigg(\lambda_{1}x^2+\lambda_{2}(1-x)^2\nonumber\\&+\lambda_{4}x(1-x)
+\lambda_{7}\xi_{1}^2x(1-x)\bigg)+\lambda_{3}(1-y)^2\nonumber\\&+ y(1-y)\bigg(
\lambda_{5}x+\lambda_{6}(1-x)+\lambda_{8}\xi^2_{2}x+\lambda_{9}\xi_{3}^2(1-x)\bigg).
\end{align}
The expression \eqref{eq:Mr} has the following form:   
\begin{align}\label{eq:Mss}
f(\chi)=a\chi^2+b(1-\chi)^2+c\chi(1-\chi),
\end{align}
The copositivity of the expression \eqref{eq:Mss} leads to \cite{ref2}:
\begin{align}\label{eq:Mrpapa}
a>0,~~ b>0~~ \text{and}~~c+2\sqrt{ab}>0.
\end{align}
Applying this criterion on \eqref{eq:Mr}, we get the following conditions:
\begin{align}
A&\equiv\lambda_{1}x^2+\lambda_{2}(1-x)^2+\lambda_{4}x(1-x)+\lambda_{7}\xi_{1}^2x(1-x)>0,  \label{eq:A}\\
B&\equiv \lambda_{3}>0 \label{eq:B}\\
C&\equiv\lambda_{5}x+\lambda_{6}(1-x)+\lambda_{8}\xi_{2}^2x+\lambda_{9}\xi_{3}^2(1-x)+2\sqrt{AB}>0 \label{eq:C}. 
\end{align}
From the expression \eqref{eq:A}, we find:
\begin{align}
\lambda_{1}&>0,\qquad \lambda_{2}>0, \nonumber\\ \lambda_{4}&+2\sqrt{\lambda_{1}\lambda_{2}}>0\nonumber.\\ \lambda_{4}&+\lambda_{7}+2\sqrt{\lambda_{1}\lambda_{2}}>0\label{eq:delenda6} .
\end{align}    
While the expression \eqref{eq:B} leads to:
\begin{align}\label{eq:delenda66}
 \lambda_{3}>0,
\end{align}
From the expression \eqref{eq:C}, we distinguish two cases:\\
If $\lambda_{6}~\text{and}~\lambda_{7}>0$, one get \eqref{eq:delenda6}-\eqref{eq:delenda66},
while, if $\lambda_{5}~\text{or}~\lambda_{6}<0$, we obtain the following:
\begin{align}
-2\sqrt{\lambda_{1}\lambda_{3}}&<\lambda_{5}< 2\sqrt{\lambda_{1}\lambda_{3}},\nonumber\\
-2\sqrt{\lambda_{2}\lambda_{3}}&<\lambda_{6}< 2\sqrt{\lambda_{2}\lambda_{3}},\nonumber\\
-2\sqrt{\lambda_{1}\lambda_{3}} &<\lambda_{5}+\lambda_{8}< 2\sqrt{\lambda_{1}\lambda_{3}},\nonumber\\
-2\sqrt{\lambda_{2}\lambda_{3}}&<\lambda_{6}+\lambda_{9}< 2\sqrt{\lambda_{2}\lambda_{3}}\nonumber,\\
4(\lambda_{4}+\lambda_{7})\lambda_{3}-&2(\lambda_{5}+\lambda_{8})(\lambda_{6}+\lambda_{9})+2\sqrt{\Lambda_{1}}>0,\nonumber\\
4\lambda_{4}\lambda_{3}-2\lambda_{5}\lambda_{6}&+2\sqrt{\Lambda_{2}}>0,\nonumber\\
4\lambda_{4}\lambda_{3}-2\lambda_{5}(\lambda_{6}&+\lambda_{9})+2\sqrt{\Lambda_{3}}>0,\nonumber\\
4\lambda_{4}\lambda_{3}-2(\lambda_{5}&+\lambda_{8})(\lambda_{6}+\lambda_{9})+2\sqrt{\Lambda_{4}}>0,\nonumber\\
4\lambda_{4}\lambda_{3}-2(\lambda_{5}&+\lambda_{8})\lambda_{6}+2\sqrt{\Lambda_{5}}>0,\nonumber\\
4(\lambda_{4}+\lambda_{7})\lambda_{3}&-2\lambda_{5}\lambda_{6}+2\sqrt{\Lambda_{6}}>0,\nonumber\\
4(\lambda_{4}+\lambda_{7})\lambda_{3}&-2\lambda_{5}(\lambda_{6}+\lambda_{9})+2\sqrt{\Lambda_{7}}>0,\nonumber\\
4(\lambda_{4}+\lambda_{7})\lambda_{3}&-2(\lambda_{5}+\lambda_{8})\lambda_{6}+2\sqrt{\Lambda_{8}}>0\label{eq:delenda2}.
\end{align}
Where:
\begin{align}
\Lambda_{1}&=(4\lambda_{1}\lambda_{3}-(\lambda_{5}+\lambda_{8})^2)(4\lambda_{2}\lambda_{3}-(\lambda_{6}+\lambda_{9})^2),\\
\Lambda_{2}&=(4\lambda_{1}\lambda_{3}-\lambda_{5}^2)(4\lambda_{2}\lambda_{3}-\lambda_{6}^2,\\
\Lambda_{3}&=(4\lambda_{1}\lambda_{3}-\lambda_{5}^2)(4\lambda_{2}\lambda_{3}-(\lambda_{6}+\lambda_{9})^2),\\
\Lambda_{4}&=(4\lambda_{1}\lambda_{3}-(\lambda_{5}+\lambda_{8})^2)(4\lambda_{2}\lambda_{3}-(\lambda_{6}+\lambda_{9})^2),\\
\Lambda_{5}&=(4\lambda_{1}\lambda_{3}-(\lambda_{5}+\lambda_{8})^2)(4\lambda_{2}\lambda_{3}-\lambda_{6}^2),\\
\Lambda_{6}&=(4\lambda_{1}\lambda_{3}-\lambda_{5}^2)(4\lambda_{2}\lambda_{3}-\lambda_{6}^2),\\
\Lambda_{7}&=(4\lambda_{1}\lambda_{3}-\lambda_{5}^2)(4\lambda_{2}\lambda_{3}-(\lambda_{6}+\lambda_{9})^2),\\
\Lambda_{8}&=(4\lambda_{1}\lambda_{3}-(\lambda_{5}+\lambda_{8})^2)(4\lambda_{2}\lambda_{3}-\lambda_{6}^2).
\end{align}
\indent All the previous conditions \eqref{eq:delenda6}-\eqref{eq:delenda2} ensure the necessary and sufficient conditions for the boundedness of the scalar potential from below in any direction in the field space. Together with the minimization conditions resulted from the positivity of the Hessian matrix \eqref{eq:delenda1}, we determine the first set of the theoretical constraints. 
\subsection{Perturbative Unitarity bounds and the positivity of the scalar bosons masses}
\label{sec:Perturbative Unitarity bounds and scalar mass spectrum}
\indent Other constraints on the scalar potential parameters are
obtained from the unitarity conditions. In order to derive those constraints one needs to look at the tree level scattering processes: scalar-scalar
scattering, gauge boson–gauge boson scattering, and
scalar–gauge boson scattering \cite{ref15}. \\
 \indent Applying the equivalence theorem \cite{ref17, ref2}.
 The unitarity constraints at the tree level in the compact 341 model can be implemented by considering only scalar-scalar scattering processes dominated by quartic interactions.\\
\indent The perturbative unitarity conditions are obtained in many BSM models \cite{ref2, ref15, ref16} by using the S matrix for all the elastic scatterings of two body scalar boson states, even in the SM this idea has been used to constrain the theoretical limits over the SM Higgs boson mass.\\
\indent The scattering amplitude for any $2\longrightarrow 2$ process can be expressed in terms of the Legendre polynomial and the partial wave amplitude $a_{J}$ \cite{ref1706,ref68,ref69} where $a_{J}$ can be expressed as:
\begin{align}\label{eq:nb}
\text{Im}(a_{J})=|a_{J}|^{2}.
\end{align}
The expression \eqref{eq:nb} is an equation of a circle with the radius $\frac{1}{2}$ and a center (0, $\frac{1}{2}$). In the high energy limit, it can be shown that the unitarity condition requires:
\begin{align}\label{eq:lkvr}
|Re(a_{0})|<\frac{1}{2}
\end{align}
\indent In general, the constrain \eqref{eq:lkvr} constraints the scattering amplitude for all possible two particle states $S_{1}S_{2}\longrightarrow S_{ 3}S_{4}$ processes as follows:
\begin{align}\label{eq:lk}
|\mathcal{M}|<8 \pi
\end{align}
Where the $S_{i}$ (i=1,..4) represent all (pseudo) scalar bosons in the model. The unitarity constraint is found by applying the bound \eqref{eq:lk} on all possible eigenvalues of all scattering matrices.\\
\indent The compact 341 model contains many scalar fields components (three quadruplets scalar fields), therefore, it is difficult to calculate all the eigenvalues of all possible matrices for all the elastic scatterings of two body scalar boson states. Fortunately, there is an alternative method used in ref \cite{ref17}. Instead of extracting the S-Matrices and calculate all the eigenvalues, we derive all possible quartic contact terms as a function of the physical scalar fields \cite{ref17}. In this way we can immediately find out the unitarity bounds on the quartic couplings.\\
\indent In the appendix, we list the possible non-zero quartic couplings that appear in the compact 341 after expanding the full scalar potential in terms of the physical quartic couplings. Using the fact that $\upsilon_{\eta}$=$\upsilon_{\chi}$, $\upsilon_{\rho}\ll\upsilon_{\eta}$ and $\upsilon_{\rho}\ll\upsilon_{\chi}$, we find $C_{\alpha}^2=1$, $S^2_{\alpha}=0$, $C_{\beta}^2=S_{\beta}^2=\frac{1}{2}$ and all terms with quartic couplings in four scalar scattering are bounded by $8\pi$. Then the unitarity constraints on the scalar parameters are:
\begin{align}
\alpha^2\lambda_{4}&+\gamma^2(\lambda_{6}+\lambda_{9})<16\pi,\nonumber\\
\beta^2\lambda_{4}&+\sigma^2(\lambda_{6}+\lambda_{9})<16\pi,\nonumber\\
\alpha\beta\lambda_{4}&+\gamma\sigma(\lambda_{6}+\lambda_{9})<8\pi,\nonumber\\
(\alpha^2&+2\alpha\beta)\lambda_{4}+\gamma^2\lambda_{6}<32\pi,\nonumber\\
\lambda_{4}&+\lambda_{6}<32\pi,\nonumber\\
\beta^2\lambda_{4}&+\sigma^2\lambda_{6}<32\pi,\nonumber\\
\sigma\lambda_{9}&+\beta\lambda_{7}<16\pi,\nonumber\\
\gamma\lambda_{9}&+\alpha\lambda_{7}<16\pi,\nonumber\\
\gamma\sigma\lambda_{6}&<16\pi,\nonumber\\
\alpha^2\lambda_{4}&+\gamma^2\lambda_{6}+\alpha^2\lambda_{7}<16\pi,\nonumber\\
\beta^2\lambda_{4}&+\sigma^2\lambda_{6}+\beta^2\lambda_{7}<16\pi,\nonumber\\
\alpha\beta\lambda_{4}&+\sigma\gamma\lambda_{6}+\alpha\beta\lambda_{7}<8\pi,\nonumber\\
2\alpha^2\lambda_{1}&+2\gamma^2\lambda_{3}+\lambda_{5}(\alpha^2+\gamma^2)+\lambda_{8}(\alpha^2+\gamma^2\nonumber\\&+(\sqrt{2}+1)\gamma\alpha)<32\pi,\nonumber\\
2\beta^2\lambda_{1}&+2\sigma^2\lambda_{3}+\lambda_{5}(\beta^2+\sigma^2)+\lambda_{8}(\beta^2+\sigma^2\nonumber\\&+\sigma\beta)<32\pi,\nonumber\\
2\beta\alpha\lambda_{1}&+4\sigma\gamma\lambda_{3}+2\lambda_{5}(\beta\alpha+\gamma^2)+\lambda_{8}(2\beta\alpha+2\sigma\alpha\nonumber\\&+\beta+\sigma\gamma+\beta\gamma+\sigma\beta)<32\pi\nonumber,\\
\lambda_{1}&+\lambda_{3}+\lambda_{5}<32\pi,\nonumber\\
\alpha^4\lambda_{1}&+\gamma^2\lambda_{3}<32\pi,\nonumber\\
\beta^4\lambda_{1}&+\sigma^4\lambda_{3}+\sigma^2\beta^2<32\pi,\nonumber\\
6\beta^2\alpha^2\lambda_{1}&+6\sigma^2\gamma^2\lambda_{3}+\lambda_{5}(4\alpha\beta\gamma\sigma+\gamma^2\alpha^2+\gamma^2\beta^2\nonumber\\&+\alpha^2\sigma^2)<32\pi,\nonumber\\
2\alpha^3\beta\lambda_{1}&+2\lambda_{3}\gamma^3\sigma+\lambda_{5}(\beta\alpha\gamma^2+\alpha^2\gamma\sigma)<16\pi,\nonumber\\
2\beta^3\alpha\lambda_{1}&+2\lambda_{3}\sigma^3\gamma+\lambda_{5}(\beta\alpha\sigma^2+\beta^2\gamma\sigma)<16\pi,\nonumber\\
\lambda_{4}&+\lambda_{6}+\lambda_{9}<16\pi,\nonumber\\
\lambda_{4}&+\lambda_{6}+\lambda_{7}<16\pi.
\end{align}
Where $\alpha$, $\beta$, $\gamma$ and $\sigma$ have the following expressions:
\begin{align}
\alpha&=\frac{-\sqrt{X^{2}+(Y-\sqrt{X^{2}+Y^{2}})^{2}}}{\sqrt{4(X^{2}+Y^{2})}},\\
\beta&=\frac{\sqrt{X^{2}+(Y+\sqrt{X^{2}+Y^{2}})^{2}}}{\sqrt{4(X^{2}+Y^{2})}},
\end{align}
\begin{align}
\gamma&=\frac{(Y+\sqrt{X^{2}+Y^{2}})(\sqrt{X^{2}+(Y-\sqrt{X^{2}+Y^{2}})^{2}})}{X\sqrt{4(X^{2}+Y^{2})}},\\
\sigma&=\frac{-(Y-\sqrt{X^{2}+Y^{2}})(\sqrt{X^{2}+(Y+\sqrt{X^{2}+Y^{2}})^{2}})}{X\sqrt{4(X^{2}+Y^{2})}}.\\
\text{With}~~X&=\lambda_{5},\qquad Y=\lambda_{1}-\lambda_{3}.
\end{align}
Moreover, to maintain the perturbativity of the model, all the quartic couplings of the scalar potential $\lambda_i(i=1...9)$ must satisfy this condition:
\begin{align}
|\lambda_{i}|\leq4\pi
\end{align}
 \indent In the compact 341 model, the physical scalar bosons masses are fully determined by the parameters of the scalar potential $\lambda_{i}$, therefore, other constraints on the scalar parameters can be found from the positivity of all scalar bosons masses. As we reported previously, the physical spectrum consists of three CP-even scalars, $H_{i}$ (i=1,2,3) 
four charged scalars, $H_{i}^{\mp}$ (i=1,2) and two doubly charged scalar bosons $H^{\mp\mp}$. A complementary set of constraints on the $\lambda_{i}$ comes from the positivity of all the masses of the scalar bosons is:
\begin{align}
\lambda_{1}&+\lambda_{3}-\sqrt{(\lambda_{1}-\lambda_{3})^{2}+\lambda_{5}^{2}}>0,\nonumber\\
\lambda_{1}&+\lambda_{3}+\sqrt{(\lambda_{1}-\lambda_{3})^{2}+\lambda_{5}^{2}}>0,\nonumber\\
\lambda_{2}&+\frac{\lambda_{3}\lambda^{2}_{4}+\lambda_{6}(\lambda_{1}\lambda_{6}-\lambda_{4}\lambda_{5})}{\lambda^{2}_{5}-4\lambda_{1}\lambda_{3}}>0,\nonumber\\
\lambda_{7}&>0,\qquad \lambda_{8}>0,\qquad \lambda_{9}>0.
\end{align}
 \indent In addition, the scalar parameters are constrained by another strong condition by imposing that the lightest scalar boson $H_{1}$ is identical to the Standard Model Higgs like boson, by talking $M_{H_{1}}$= 125.09 GeV and $\upsilon_{\rho}$=246 GeV, therefore:
\begin{align}
\lambda_{2}+\frac{\lambda_{3}\lambda^{2}_{4}+\lambda_{6}(\lambda_{1}\lambda_{6}-\lambda_{4}\lambda_{5})}{\lambda^{2}_{5}-4\lambda_{1}\lambda_{3}}=\frac{m^{2}_{h_{1}}}{\upsilon^{2}_{\rho}},
\end{align}
 \indent Moreover, the ref \cite{ref1} reported that the compact 341 model has a landau pole $\Lambda$ around 5 TeV, that leads to a stringent constraint on the parameters. It requires that all scalar bosons masses and all VEVs are bounded to be less or equal $\Lambda$.\\
 \indent We give random numbers for $\lambda_{i}$ (i=1..9) taking into account all the theoretical constraints. We choose the following benchmark point to study the variation of the scalar bosons masses as a function of the vacuum expectation value $\upsilon_{\eta}$. 
\begin{align}
\lambda_{1}&, \lambda_{2}, \lambda_{3}, \lambda_{4}, \lambda_{5} \lambda_{6}, \lambda_{7}, \lambda_{8}, \lambda_{9}\equiv(1.24916, 1.4595, 2.08534, \nonumber\\& 0.612214, 0.544161, -2.88788, 1.20945, 
0.308258, 3.64476)\nonumber.
\end{align}
 \indent The allowed region is smaller than 5 TeV due to the exclusion limits resulted from the existence of the Landau pole (larger masses should be smaller or equal to 5 TeV).
\begin{figure}[H]
\centering
\includegraphics[width=0.45\textwidth]{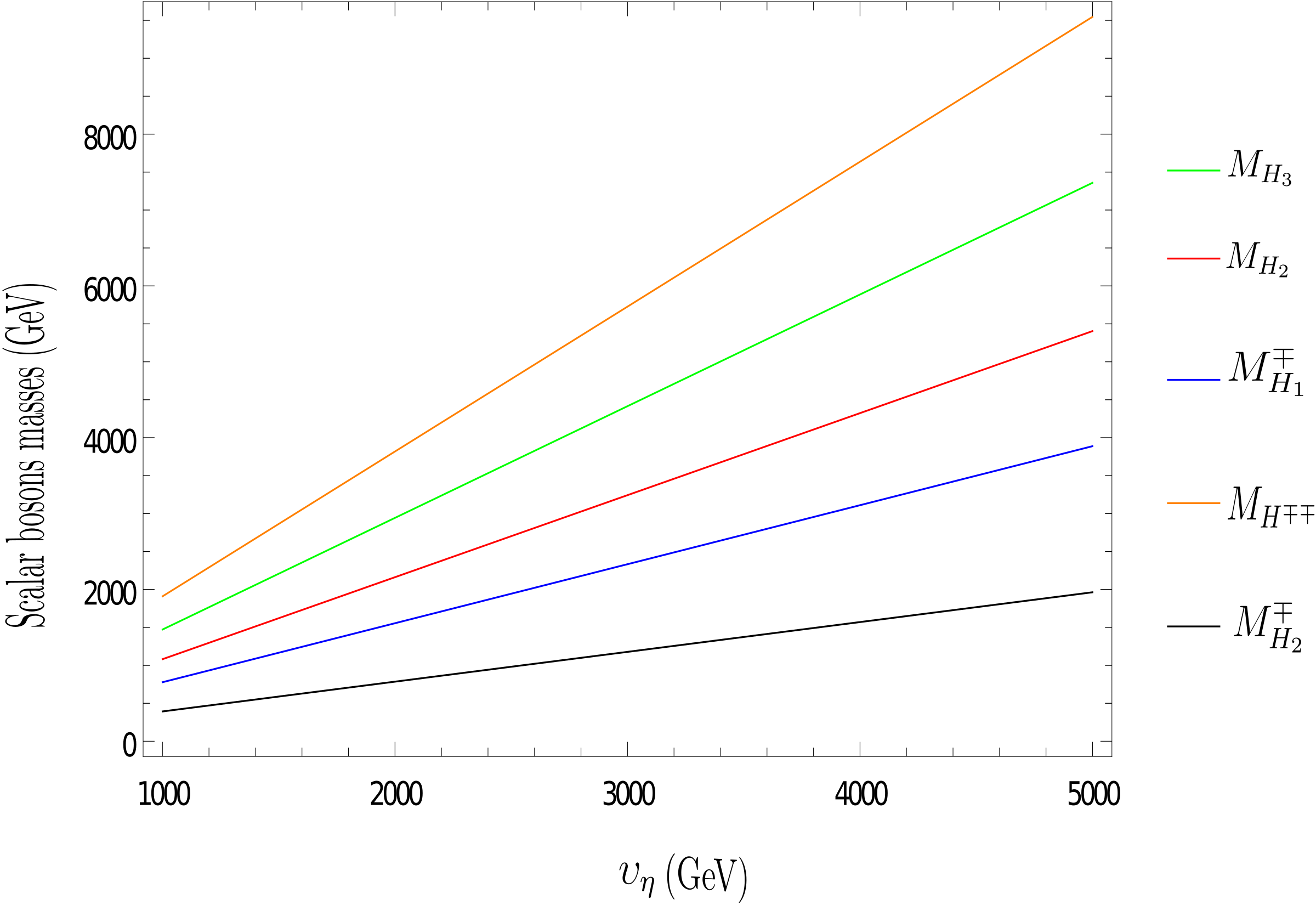}
\caption{\label{fig:leepp} The variation of the scalar bosons masses in the compact 341 model as a function of $\upsilon_{\eta}$.}
\end{figure}
\begin{figure}[H]
	\centering
	\includegraphics[width=0.48\textwidth]{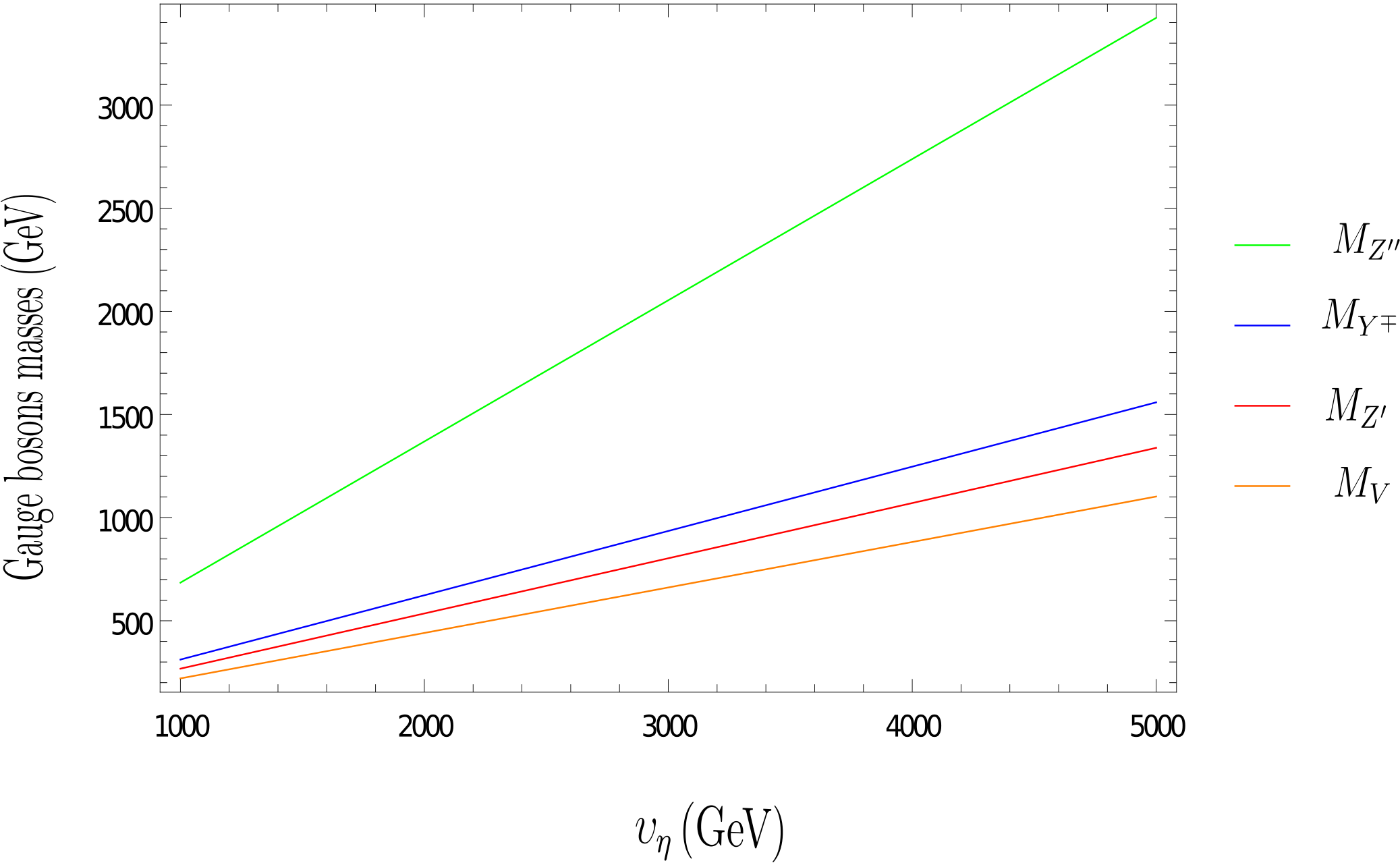}
	\caption{\label{fig:lee} The variation of the gauge bosons masses in the compact 341 model as a function of $\upsilon_{\eta}$ where V represents $K^{\prime0}$, $K^{0}$, $V^{\mp\mp}$, $X^{\mp}$ and $K_{1}^{\mp}$.}
\end{figure}
 \indent Figures \eqref{fig:leepp} and \eqref{fig:lee} represent the variation of the scalar and gauge bosons masses as a function of the $\upsilon_{\eta}$
respectively. Among all of them, $H^{\mp\mp}$ and $Z^{\prime\prime}$ are the heaviest for all the choices of $\upsilon_{\eta}$.
\section{Conclusion}
\label{sec:Conclusion}
 \indent The theoretical constraints on the parameters space are an important issue that must be imposed on the scalar potential couplings in any theory beyond the Standard Model. Thus
one needs to find the  necessary conditions to ensure the allowed region for the scalar couplings.\\
 \indent In our work, the corresponding theoretical constraints in the compact 341 model are derived such as the vacuum stability, minimization of the scalar potential, perturbative unitarity bounds and perturbativity of the scalar potential couplings. Moreover, other conditions coming from the positivity of the scalar bosons masses with the stringent condition of the landau pole are also derived.\\
 \indent We have used parametrizations which allows us to find analytically the conditions which guarantee the boundedness of the scalar potential in all the directions. Together with the positivity of the Hessian matrix resulted from the first and second derivative of the scalar potential, we derive the first set of the theoretical constraints on the scalar couplings.
The second objective of this paper is to derive the tree level conditions for the quartic couplings of the scalar potential coming from the perturbative unitarity conditions,  we express the quartic couplings in terms of the physical scalar fields instead of calculating the s-wave amplitude matrix
for all possible 2 to 2 body (pseudo) scalar boson elastic scatterings in the high energy limit. Also, the positivity of all scalar bosons masses are taken into account imposing additional constraints on  $\lambda_{i}$ (i=1..9). Finally, We have used the fact that all quartic scalar couplings are smaller than 4$\pi$ to ensure the perturbativity of the scalar potential. \\
\indent The combination of all those
theoretical constraints  together with the emergence of the Landau pole
at around 5 TeV determine the allowed regions of the parameters space which must be taken
into account in our future phenomenological studies (work in progress) in the context of the compact 341 model.
Moreover, we have presented the benchmark point which we suggest to study the variation of the scalar bosons masses as a function of the $\upsilon_{\eta}$.
\acknowledgments
	We  would like to thank Prof Dominik Stöckinger for many helpful discussions and clarifying comments. We also thank prof C. A. de S. Pires, Garv Chauhan and Uladzimir Khasianevich for useful discussions. We are very grateful to the Algerian ministry of higher education and scientific research and DGRSDT for the financial support. 
	\appendix*
	\section{Scalar Quartic Couplings}
	\label{sec:app:Scalar Quartic Couplings}
		All possible non-zero quartic couplings in terms of the physical scalar fields that appear in the compact 341 model are:
	\begin{align}
	H_{1}H_{1}H_{1}H_{1}&: \frac{\lambda_{2}}{4} \nonumber,\\
	H^{--}H^{++}H_{2}H_{2}&:\lambda_{3}S_{\alpha}^{2}\gamma^{2}+ \lambda_{4}\frac{\alpha^{2}C_{\alpha}^{2}}{2}+(\lambda_{6}+\lambda_{9})\frac{\gamma^{2}C_{\alpha}^{2}}{2}\nonumber\\&-\lambda_{5}\frac{\alpha^{2}S_{\alpha}^{2}}{2}\nonumber,\\
	H^{--}H^{++}H_{3}H_{3}&:\lambda_{3}S_{\alpha}^{2}\sigma^{2}+\lambda_{4}\frac{C^{2}_{\alpha}\beta^{2}}{2}+(\lambda_{6}+\lambda_{9})\frac{C^{2}_{\alpha}\sigma^{2}}{2}
\nonumber\\&-\lambda_{5}\frac{\beta^{2}S_{\alpha}^{2}}{2}\nonumber,\\
	H^{--}H^{++}H_{2}H_{3}&:2\lambda_{3}\sigma\gamma S_{\alpha}^{2}+C^{2}_{\alpha}\alpha\beta\lambda_{4}+\sigma\gamma C_{\alpha}^{2}(\lambda_{6}+\nonumber\\&\lambda_{9})
	\alpha\beta S_{\alpha}^{2}\lambda_{5}\nonumber,
		\end{align}
	\begin{align}
	H^{-}_{2}H^{+}_{2}H_{1}H_{1}&:\lambda_{6}\frac{C_{\beta}^{2}}{2}+\lambda_{4}\frac{S_{\beta}^{2}}{2}\nonumber,\\
	H_{2}H_{2}H_{1}H_{1}&:\lambda_{4}(\frac{\alpha^{2}}{4}+\frac{\alpha\beta}{2})+\lambda_{6}\frac{\gamma^{2}}{4}\nonumber,\\
	H_{3}H_{3}H_{1}H_{1}&:\lambda_{6}\frac{\sigma^{2}}{4}+\lambda_{4}\frac{\beta^{2}}{4}\nonumber,\\
	H^{--}H_{2}^{+}H^{+}_{1}H_{3}&:\lambda_{8}(\frac{\beta}{\sqrt{2}}C_{\beta}S_{\alpha}^{2}+\frac{\sigma}{\sqrt{2}}S_{\beta}S_{\alpha}^{2})+\lambda_{9}\frac{C^{2}_{\alpha}S_{\beta}\sigma}{\sqrt{2}}\nonumber\\
	&+\lambda_{7}\frac{C^{2}_{\alpha}S_{\beta}\beta}{\sqrt{2}}\nonumber,\\
	H^{--}H_{2}^{+}H^{+}_{1}H_{1}&:\lambda_{9}\frac{C_{\beta}C_{\alpha}S_{\alpha}}{\sqrt{2}}+\lambda_{7}\frac{S_{\alpha}C_{\alpha}S_{\beta}}{\sqrt{2}},\nonumber\\
	H^{-}_{1}H_{2}^{-}H^{++}_{1}H_{2}&:\lambda_{9}\frac{\gamma}{\sqrt{2}}C_{\alpha}^{2}C_{\beta}+\lambda_{7}\frac{\alpha}{\sqrt{2}}C_{\alpha}^{2}S_{\beta}+\lambda_{8}\frac{\alpha}{\sqrt{2}}S_{\alpha}^{2}C_{\beta}\nonumber,\\
	H^{-}_{1}H_{2}^{-}H^{++}_{1}H_{3}&:\lambda_{9}\frac{\sigma}{\sqrt{2}}C_{\alpha}^{2}C_{\beta}+\lambda_{7}\frac{\beta}{\sqrt{2}}C_{\alpha}^{2}S_{\beta}\nonumber,\\
	H^{--}H^{++}H_{1}H_{2}&:\lambda_{9}\dfrac{\gamma}{2}C_{\alpha}S_{\alpha},\nonumber\\
	H^{--}H^{++}H_{1}H_{3}&:\lambda_{9}\dfrac{\sigma}{2}C_{\alpha}S_{\alpha},\nonumber\\
	H^{++}H^{-}_{1}H^{-}_{2}H_{1}&:\lambda_{7}\dfrac{\sigma}{2}S_{\beta}C_{\alpha}S_{\alpha},\nonumber\\
	H_1^{+}H_1^{-}H_1H_3&: \lambda_{7}C_{\alpha}S_{\alpha}\beta,\nonumber\\
	H_{1}H_{1}H_{2}H_{3}&:\frac{\lambda_{6}\gamma\sigma}{2},\nonumber\\
	H_{1}^{+}H_{1}^{-}H_{1}H_{2}&:\lambda_{7}\frac{\alpha}{2}C_{\alpha}S_{\alpha},\nonumber\\
	H_{1}^{+}H_{1}^{-}H_{1}^{+}H_{1}^{-}&:\lambda_{1}S_{\alpha}^{4}+\lambda_{2}C_{\alpha}^{4}+\lambda_{4}S_{\alpha}^{2}C_{\alpha}^{2}\nonumber,\\
	H^{++}H_1^{-}H_1^-H_2&: \frac{\lambda_{8}}{\sqrt{2}}S^2_{\alpha}S_{\beta}\gamma, \nonumber\\ H^{++}H^{--}H^{++}H^{--}&: \lambda_{2}C^{2}_{\alpha}+\lambda_{3}S_{\alpha}^4+\lambda_{6}C^{2}_{\alpha}S^{2}_{\alpha},\nonumber\\
	H_{1}^{+}H_{1}^{-}H_{2}H_{2}&:\lambda_{1}S_{\alpha}^{4}\alpha^{2}+\lambda_{4}\frac{\alpha^{2}}{2}C_{\alpha}^{2}+\lambda_{5}\frac{\gamma^{2}}{2}S_{\alpha}^{2}
	\nonumber\\&+\lambda_{6}\frac{\gamma^{2}}{2}C_{\alpha}^{2}+\lambda_{7}\frac{\alpha^{2}}{2}C_{\alpha}^{2}\nonumber,\\
	H_{1}^{+}H_{1}^{-}H_{3}H_{3}&:\lambda_{1}S_{\alpha}^{2}\beta^{2}+\lambda_{4}\frac{\beta^{2}}{2}C_{\alpha}^{2}+\lambda_{5}\frac{\sigma^{2}}{2}S_{\alpha}^{2}+\lambda_{6}\frac{\sigma^{2}}{2}C_{\alpha}^{2}\nonumber\\&+\lambda_{7}\frac{\beta^{2}}{2}C_{\alpha}^{2}\nonumber,\\
	H_{1}^{+}H_{1}^{-}H_{2}H_{3}&:\lambda_{1}S_{\alpha}^{2}\beta\alpha+\lambda_{4}\beta\alpha C_{\alpha}^{2}+\lambda_{5}\sigma\gamma S_{\alpha}^{2}+\lambda_{6}\sigma\gamma C_{\alpha}^{2}\nonumber\\&+\lambda_{7}\beta\alpha C_{\alpha}^{2}\nonumber,\\
		H^{++}H_1^{-}H_1^-H_3&: \frac{\lambda_{8}}{\sqrt{2}}(\beta S^2_{\alpha}C_{\beta}+\gamma S^2_{\alpha}S_{\beta})\nonumber,\\
	H_{2}^{+}H_{2}^{-}H_{2}H_{2}&:\lambda_{1}S_{\beta}^{2}\alpha^2+\lambda_{3}\gamma^2C_{\beta}^{2}+\lambda_{5}(\frac{\alpha^2}{2} C_{\beta}^{2}+\dfrac{\gamma^2}{2} S_{\beta}^{2})\nonumber\\& +\lambda_{8}(\frac{\alpha^2}{2}C_{\beta}^{2}+\frac{\gamma\alpha}{2}C_{\alpha}S_{\beta}+\frac{\gamma^2}{2}S_{\beta}^{2}+\frac{\gamma\alpha}{2}C_{\beta}S_{\beta})\nonumber,\\
	H_{2}H_{2}H_{2}H_{3}&: \lambda_{1}\alpha^3\beta+\lambda_{3}\gamma^3\sigma+\frac{\lambda_{5}}{2}(\alpha\beta\gamma^2+\alpha^2\gamma\sigma),\nonumber\\
	H_{2}^{+}H_{2}^{-}H_{3}H_{3}&:\lambda_{1}S_{\beta}^{2}\beta^2+\lambda_{3}\sigma^2C_{\beta}^{2}+\lambda_{5}(\frac{\beta^2}{2} C_{\beta}^{2}+\frac{\sigma^2}{2} S_{\beta}^{2}) \nonumber\\&+\frac{\lambda_{8}}{2}(\beta^2C_{\beta}^{2}+\sigma^2S_{\beta}^{2}+\sigma\beta C_{\beta}S_{\beta})\nonumber,
		\end{align}
	\begin{align}
	H_{2}^{+}H_{2}^{-}H_{2}H_{3}&:\lambda_{1}S_{\beta}^{2}\beta\alpha+2\lambda_{3}\sigma\gamma C_{\beta}^{2}+\lambda_{5}(\beta\alpha C_{\beta}^{2}+\sigma\gamma S_{\beta}^{2})\nonumber\\& +\frac{\lambda_{8}}{2}(2\beta\alpha C_{\beta}^{2}+(\alpha\sigma+\beta)S_{\beta}C_{\beta}\nonumber\\&+\sigma\gamma S^2_{\beta}
	+(\alpha\sigma+\beta\gamma)C_{\beta}S_{\beta}+\sigma\beta C_{\beta}S_{\beta}),\nonumber\\
	H_{2}^{+}H_{2}^{-}H_{2}^{+}H_{2}^{-}&: S_{\beta}^{4}\lambda_{1}+\lambda_{3}C_{\beta}^{4}+\lambda_{5}C_{\beta}^{2}S_{\beta}^{2},\nonumber\\
	H_{2}H_{2}H_{2}H_{2}&: \lambda_{1}\frac{\alpha^4}{4}+\lambda_{3}\frac{\gamma^2}{4},\nonumber\\
	H_{3}H_{3}H_{3}H_{3}&: \lambda_{1}\frac{\beta^4}{4}+\lambda_{3}\frac{\sigma^4}{4}+\lambda_{5}\frac{\sigma^2\beta^2}{4},\nonumber\\
	H_{2}H_{3}H_{3}H_{3}&: \lambda_{1}\alpha\beta^3+\lambda_{3}\gamma\sigma^3+\frac{\lambda_{5}}{2}(\alpha\beta\sigma^2+\beta^2\gamma\sigma),\nonumber\\
	H_{1}^+H_{1}^-H_{1}H_{1}&: \lambda_{2}C^{2}_{\alpha}+\frac{\lambda_{4}}{2}S^{2}_{\alpha}+\frac{\lambda_{7}}{2}S^{2}_{\alpha},\nonumber\\
	H^{++}H^{--}H_{1}^+H_{1}^-&:2 \lambda_{2}C^{2}_{\alpha}+\lambda_{4}S_{\alpha}^4+\lambda_{5}S^{4}_{\alpha}+\lambda_{6}C^{2}_{\alpha}S^{2}_{\alpha}+\lambda_{8}S^{4}_{\alpha},\nonumber
		\end{align}
	\begin{align}
	H^{++}H^{--}H_2^{+}H_2^{-}&:2 \lambda_{3}S^{2}_{\alpha}C^{2}_{\beta}+\lambda_{4}S_{\beta}^2C^{2}_{\alpha}+\lambda_{5}S^{2}_{\alpha}S^{2}_{\beta}+
	\lambda_{6}C^{2}_{\alpha}C^{2}_{\beta}\nonumber\\&+\lambda_{7}C^{2}_{\alpha}S^{2}_{\beta},\nonumber\\
	H^{+}_{1}H_{2}^{+}H^{--}H_{2}&:\lambda_{9}\frac{\gamma}{\sqrt{2}}C_{\alpha}^{2}C_{\beta}+\lambda_{7}\frac{\alpha}{\sqrt{2}}C_{\alpha}^{2}S_{\beta}+\lambda_{8}(\frac{\gamma}{\sqrt{2}}S_{\beta}S_{\alpha}^{2}\nonumber\\&+\frac{\alpha
	}{\sqrt{2}}C_{\beta}S_{\alpha}^{2}),\nonumber\\
	H_1^{+}H_1^{-}H_1H_2&: \frac{\lambda_{7}}{2}C_{\alpha}S_{\alpha}\alpha,\nonumber\\
	H_{1}^+H_{1}^-H_{2}^-H_{2}^+&: 2\lambda_{1}S_{\beta}^2S_{\alpha}^{2}+\lambda_{4}C_{\alpha}^2S_{\beta}^2+(\lambda_{6}+\lambda_{9})C_{\alpha}^2C_{\beta}^2
\nonumber\\	&+\lambda_{5}S_{\alpha}^{2}C_{\beta},\nonumber\\
H_{2}H_{2}H_{3}H_{3}&: \lambda_{1}\frac{3\beta^2\alpha^2}{2}+\lambda_{3}\frac{3\sigma^2\gamma^2}{2}
+\lambda_{5}(\alpha\beta\gamma\sigma+\frac{\gamma^2\alpha^2}{4}\nonumber\\&+
\frac{\gamma^2\beta^2}{4}+\frac{\sigma^2\alpha^2}{4})\nonumber.\\
H^{++}H^{--}H_{1}H_{1}&: \lambda_{2}C^{2}_{\alpha}+\frac{\lambda_{9}}{2}S_{\alpha}+\frac{\lambda_{6}}{2}S^{2}_{\alpha}.\\\nonumber
	\end{align}

\end{document}